\documentclass[aps,prb,twocolumn,floatfix]{revtex4}
\usepackage{graphicx}
\begin{document}

\title{Cryogenic small-signal conversion with relaxation oscillations in
Josephson junctions}

\author{Miha Furlan}
\affiliation{Laboratory for Astrophysics, Paul Scherrer Institute,
5232 Villigen PSI, Switzerland}

\date{\today}

\begin{abstract}
Broadband detection of small electronic signals from
cryogenic devices, with the option of simple implementation
for multiplexing, is often a highly desired, although
non-trivial task. We investigate and demonstrate a
small-signal analog-to-frequency conversion system based
on relaxation oscillations in a single Josephson junction.
Operation and stability conditions are derived, with
special emphasis on noise analysis, showing the dominant
noise sources to originate from fluctuation processes in the
junction.
At optimum conditions the circuit is found to deliver excellent
noise performance over a broad dynamic range. Our simple models
apply within the regime of classical Josephson
junction and circuit dynamics, which we confirm by experimental
results. A discussion on possible applications includes a
measurement of the response to a cryogenic radiation detector.
\end{abstract}

\maketitle


\section{Introduction\label{intro.sec}}
Cryogenic devices are widely used in a broad range of applications
like radiation detection, quantum crypto\-gra\-phy, charge
manipulation on the single-electron level, quantum Hall
effect or in basic studies of mesoscopic transport.
Measurement of the electronic properties of such devices
usually requires sophisticated readout electronics.
Detection schemes where the samples at cryogenic
temperatures are remotely connected to room temperature
electronics generally face the problem of reduced
frequency bandwidth due to the impedance of long
readout lines. In addition, the risk of noise pickup
on the lines is intrinsically increased.
Alternatively, signal readout relatively close to
the sample can be accomplished with `Superconducting 
Quantum Interference Device' (SQUID) amplifiers,
which perform very successfully in many cases but
require a delicate setup (shielding) and are usually
constrained to commercially available systems.
Amplification or impedance transformation on-chip or very
close to the sample is also possible with a `High 
Electron-Mobility Transistor' (HEMT), the
dissipation of which may, however, quickly reach an
unacceptable level.

In recent years it has been realized that probing the
electronic transport in a cryogenic device with a
radio-frequency (RF)
signal\cite{Schoelkopf1998,Segall2002,Lu2003,%
Bylander2005,Day2003,Schmidt2005}
may have considerable advantages compared to
direct signal readout, mainly due to a substantial
extension of the bandwidth. In those schemes the power
of the reflected (or transmitted) RF signal from a
properly tuned tank circuit is related to the electronic
state of the device under test. The circuit needs to be
carefully designed to minimize back-action on the
cryogenic sample. Operation at microwave frequencies
also naturally opens a potential way for frequency-domain
multiplexing.\cite{Stevenson2002,Irwin2004}

A promising 
readout scheme, which we present in this
paper, consists of an on-chip analog-to-frequency converter,
delivering a frequency signal of large amplitude which
is easily demodulated with standard room temperature
(phase-locked loop) electronics. It has the advantages
of both the direct signal readout close to the sample and
a large frequency bandwidth. Particularly, it is much
easier to accurately analyze a frequency signal than to
transmit a low-level analog signal through long readout
lines and amplify it with room temperature equipment
which typically shows inferior performance in terms
of noise with increasing temperature.
Our low-noise converter circuit is based on a hysteretic
Josephson junction exhibiting relaxation oscillations.
Related ideas using relaxation oscillations in Josephson
junction were proposed for thermometry\cite{Gerdt1979}
or direct radiation detection,\cite{Nevirkovets1998} both
relying on the temperature dependence of
quasiparticle population in the gap singularity
peak of asymmetric junctions, and for the (double)
relaxation oscillation
SQUID,\cite{Mueck1988,Gudoshnikov1989,Adelerhof1994}
which is investigated and used as a magnetometer.
In our case the circuit converts an analog current
signal into a frequency with acceptable linearity
over a broad operation bias and dynamic range.

In Sec.~\ref{princ-op.sec} we review the basic
principle of a relaxation oscillation circuit and derive
conditions for stable operation. Results from the
model are illustrated with experimental data.
A thorough noise analysis with implications for the circuit's
readout resolution is given in Sec.~\ref{noise.sec}.
An optimized low-noise configuration with numerical
estimates is considered in Sec.~\ref{numeric.sec},
followed by a discussion on possible applications in
Sec.~\ref{appl.sec}. As an example we demonstrate
the readout of a cryogenic radiation detector.
The paper concludes with Sec.~\ref{conclus.sec}.


\section{Principle of operation\label{princ-op.sec}}
We assume a Josephson device with normal resistance $R_n$,
critical current $I_c$, junction capacitance $C_j$ and
superconducting energy gaps $\Delta_1 , \Delta_2$ , where
$0 < | \Delta_1 - \Delta_2 | \ll \Delta_1 + \Delta_2 \doteq e V_g$.
It shall be connected in series with an inductance $L$ and
both shunted with a resistor $R_s$, as shown schematically
in Fig.~\ref{scheme.f}. The circuit is eventually current
biased by a large resistor $R_b \gg R_s$. A Josephson junction
with a non-vanishing difference of the energy gaps shows a
region of negative differential resistance in the
current-voltage ($IV$) characteristics.
Voltage biasing the junction in that region, where $R_s$ acts
as voltage source with $R_s \ll R_n$, its operation is
potentially unstable and the circuit can undergo relaxation
oscillations.\cite{Albegova1969,Vernon1968,Calander1982}
A relaxation oscillation cycle, which is displayed in the
$IV$ diagram of Fig.~\ref{scheme.f}, can be separated into
four phases:

\begin{figure}
\includegraphics[width=85mm]{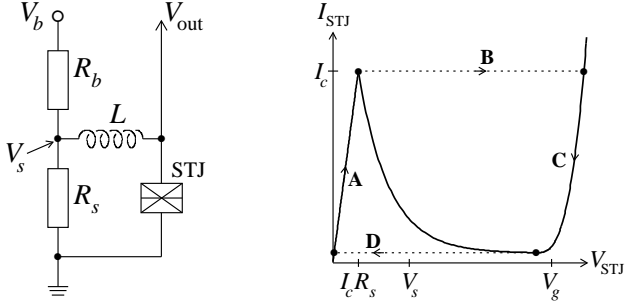}
\caption{\label{scheme.f}%
Left: Circuit diagram of the relaxation oscillator.
Right: Schematic $IV$ characteristics of a Josephson junction
with an (exaggerated) region of negative differential resistance.
The four partial processes of an oscillation cycle are labeled A,
B, C and D.}
\end{figure}

\begin{itemize}
\item (A) Initially, when $V_s = R_s I_b > R_s I_c$ is turned on,
the Josephson junction is essentially a short (supercurrent branch)
and the current through $L$ increases with a time constant
$\tau_{sc}=L/R_s$ towards a value $I_\mathrm{max} = V_s / R_s$ like
$I_{sc}(t)= I_\mathrm{max} ( 1 - \mathrm{e}^{-t / \tau_{sc}})$
until reaching $I_c$ within a time
\begin{equation}
  \tau_{_A} = -  \tau_{sc} \ln \left( 1 - \frac{R_s I_c}{V_s}\right) .
\end{equation}

\item (B) Because the junction was current biased during phase (A)
via a high-impedance $L$ it switches now to the quasiparticle branch
by developing a voltage across $C_j$ until it is charged to $V_g$.
The inductance holds the current constant if $C_j$ is small enough,
which means that
in order to observe the ``full swing'' of the voltage oscillations,
the inductive energy $I_c^2 L / 2$ and the energy from the bias
voltage $V_s^2 C_j / 2$ must be sufficient to provide the charge
on $C_j$ with $V_g$, i.e.
$I_c^2 L + V_s^2 C_j \gg (V_g-V_s)^2 C_j$
must be fulfilled.
For the case of interest where $V_s \ll V_g$ this requirement is
particularly true if
\begin{equation}
  C_j \ll \frac{I_c^2 L}{V_g^2}\, .
  \label{eq:Cmax2}
\end{equation}
Another requirement is undercritical damping of the $R_s L C_j$
circuit with
$L/(C_j R_s^2) \gg 1$.
However, because $R_s \ll V_g/I_c$, comparison with
Eq.~(\ref{eq:Cmax2}) shows that the undercritical damping
condition is already implied by (\ref{eq:Cmax2}).
The voltage switching time $\tau_{_B}$ is on the order
of $C_j V_g / I_c$ which is  negligibly short
for $C_j$ satisfying (\ref{eq:Cmax2}).

\item (C) Similarly to phase (A) the current on the quasiparticle branch
decays with $\tau_{qp}=L/R_{qp}$ (where $R_{qp}$ is the
corresponding resistance in that region of the $IV$ characteristics
{\em including} the shunt $R_s$ in series)
from $I_c$ to $I_\mathrm{min} = (V_s - V_g)/R_{qp}$ like $I_{qp}(t) =
( I_c - I_\mathrm{min}) \mathrm{e}^{-t / \tau_{qp}} + I_\mathrm{min}$
until reaching zero (or a local minimum close to zero) within time
\begin{equation}
  \tau_{_C} = \tau_{qp} \ln \left( 1+\frac{R_{qp} I_c}{V_g-V_s} \right) .
\end{equation}

\item (D) The capacitor is discharged again to zero voltage according to
the conditions in phase (B), but with a subtle difference regarding
final locking to the zero-voltage state, as discussed in the noise
section~\ref{noise_other.sec}.

\end{itemize}

Neglecting the short voltage switching times of phases (B) and (D),
the relaxation oscillation period is given by $(\tau_{_A} + \tau_{_C})$.
However, when biasing a Josephson junction
at $V_s \ll V_g$,
the oscillation dynamics are dominated by the process in phase (A) with
$\tau_{_A} \gg \tau_{_C} \approx \mathrm{O}(\tau_{_A} \frac{V_s}{V_g})$.
Therefore, the relaxation oscillation frequency is essentially given by
\begin{equation}
  f_r \approx \tau_{_A}^{-1} = - \frac{R_s}{L} \ln^{-1}
      \left( 1 - \frac{1}{\alpha} \right) .
  \label{eq:f_r_sc}
\end{equation}
where $ \alpha = I_b / I_c$ is the reduced bias current.
A series expansion for $R_{s}I_{c}\ll V_{s}\ll V_{g}$ yields
\begin{eqnarray}
  f_r & = & \frac{R_s}{L} \left\{ \alpha - \frac{1}{2} -
      \mathrm{O} \left( \frac{1}{\alpha} \right) \right\}
      \label{eq:f_r_approx1} \\
      & \approx & \frac{V_s}{I_c L}\, . \quad (\alpha \gg 1)
      \label{eq:f_r_approx2}
\end{eqnarray}
These equations describe an almost linear analog-to-frequency converter.
The same result follows from an expansion around $R_s \rightarrow 0$,
relevant for the readout of a variable resistance device in place of $R_s$.
The current-to-frequency conversion factor is
\begin{equation}
  \frac{\mathrm{d} f_r}{\mathrm{d} I_b} = \frac{R_s}{I_c L} +
      \mathrm{O}\left( \frac{1}{I_b^2} \right) .
  \label{eq:df/dI}
\end{equation}

\begin{figure}
\includegraphics[width=85mm]{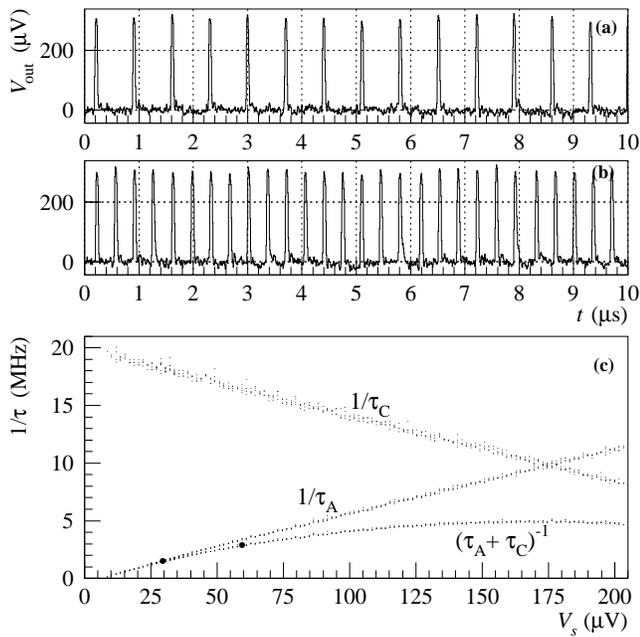}
\caption{\label{basicosc.f}%
Relaxation oscillation time-traces, plots (a) and (b),
measured at bias voltages differing by a factor of two.
Plot (c) shows
the inverse of measured time constants $\tau_{_A}$, $\tau_{_C}$
and of their sum as a function of $V_s$.
The two dots in (c) correspond to the signal traces (a) and (b).
Data were taken at $80\, \mathrm{mK}$ with an Aluminum based
Josephson junction with nominally $I_c = 58.3\, \mu\mathrm{A}$
and a circuit consisting of
$L = 280\, \mathrm{nH}$, $R_s = 91\, \mathrm{m}\Omega$.}
\end{figure}

Figure~\ref{basicosc.f} shows experimental relaxation oscillation data.
The amplitude of $V_\mathrm{out}$ corresponds to the gap voltage $V_g$,
and the dynamics follow the model predictions.
In the operation range $R_s I_c < V_s \leq 0.1 \cdot V_g$ the effective
oscillation frequency $(\tau_{_A} + \tau_{_B} )^{-1}$ deviates
from linearity by less than 10\% (a larger operation range
can also be chosen with an easy subsequent linearization of the
results according to circuit calibration).
Linear extrapolation of $\tau_{_A}^{-1}$ towards $V_s = 0$ yields a
frequency offset in agreement with Eq.~(\ref{eq:f_r_approx1}).

Relaxation oscillations in Josephson junctions can be analyzed
in terms of subharmonics of the Josephson frequency.
This implies that the number $n_{\phi}$ of Josephson
oscillations per relaxation oscillation cycle be much larger than unity
in order to prevent significant harmonic phase locking of
the two oscillating processes.
That sets a constraint on the frequency response and we can write
\begin{equation}
  f_r = \frac{V_s}{\phi_0 n_{\phi}} , \quad(n_{\phi} \gg 1)
  \label{eq:fr_max}
\end{equation}
and
\begin{equation}
  I_c L = \phi_0 n_{\phi}\, , \quad(\alpha,\: n_{\phi} \gg 1) ,
  \label{eq:IcL}
\end{equation}
where $\phi_0$ is the magnetic flux quantum.
This argument is in line with the requirement
\begin{equation}
  \beta_L = \frac{2\pi}{\phi_{0}} I_c L \gg 1
  \label{eq:beta_L}
\end{equation}
as stated in the literature.\cite{Whan1995}

\begin{figure}
\includegraphics[width=85mm]{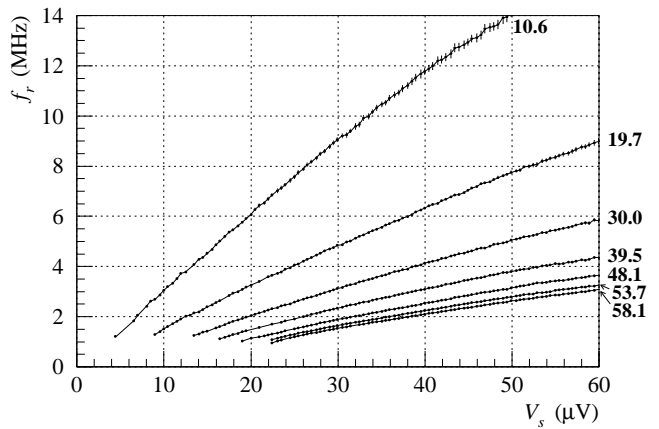}
\caption{\label{f_r(B).f}%
Measured relaxation oscillation frequencies versus bias voltage
for different $I_c$ due to application of magnetic field.
The values to the right-hand side of the curves are the effective
critical currents in $(\mu\mathrm{A})$ as obtained from fitting the
theory to the experimental data. Device and circuit parameters
are the same as in Fig.~\ref{basicosc.f}.}
\end{figure}

Modulation of  $I_c$ by application of a magnetic field parallel
to the Josephson junction results in a variation of the
relaxation oscillation frequency according to
Eq.~(\ref{eq:f_r_approx2}). This offers a convenient way to
tune the circuit's dynamic properties as well as to
extend the operation range to lower $I_b$.
Figure~\ref{f_r(B).f} shows measurements of $f_r$
as a function of $V_s$ for different $I_c$.
In order to take an $I_c$ modulation into account we denote
$\kappa = I_c^0 / I_c$ as the factor by which the nominal
value $I_c^0$ may be suppressed. In the limit $T \rightarrow 0$
and equal superconductors with gap $\Delta$ the zero-field
critical current\cite{Ambegaokar1963} is
$I_c^0 = \pi \Delta / (2 e R_n)$. Corrections due to
small gap differences are safely neglected for our analysis
and we can write
\begin{equation}
  I_c R_n \approx \frac{\pi V_g} {4 \kappa}\, .
  \label{eq:IcRn}
\end{equation}

The dimensionless capacitance parameter $\beta_c$,
which is a measure for the damping strength of the junction,
is given by
\begin{equation}
  \beta_c = (\omega_p R_n C_j)^2\, ,
  \label{eq:beta_c}
\end{equation}
where $\omega_p = (2 e I_c / \hbar C_j )^{1/2}$
is the Josephson plasma frequency.
For a weakly damped and hysteretic Josephson junction
we should choose $\beta_c$ larger than unity.
The inequality~(\ref{eq:Cmax2}) in terms of
Eqs.~(\ref{eq:IcRn},\ref{eq:beta_c}) can be rewritten as
\begin{equation}
  \beta_c \, \frac{\phi_0}{2 \pi} \left( \frac{4 \kappa}{\pi} \right)^2 \ll I_c L\, ,
  \label{eq:Lmin}
\end{equation}
and substituting $I_c L$ from (\ref{eq:IcL}) yields
\begin{equation}
  n_{\phi} \gg \beta_c \, \frac{8 \kappa^2}{\pi^3}\, .
  \label{eq:n_phi}
\end{equation}
This result is consistent with the condition $n_{\phi} \gg 1$
as we required for Eqs.~(\ref{eq:fr_max},\ref{eq:IcL}).

Finally, a comparison between (\ref{eq:beta_L}) and (\ref{eq:Lmin})
shows that the latter is the more stringent of both conditions
by the factor $\beta_c (4 \kappa / \pi)^2 > 1$.
Consequently, the very minimum of $I_c L$ is determined by relation
(\ref{eq:Lmin}), which constitutes, together with (\ref{eq:n_phi})
and $\beta_c > 1$, the relevant conditions for proper observation
of relaxation oscillations and which should help to choose
appropriate circuit components.


\section{Noise and resolution\label{noise.sec}}
\subsection{General}
In this section we list the significant current noise sources referred
to the circuit input (i.e.\ at $R_b$).
Special attention is paid to experimental mean fluctuations
of the relaxation oscillation periods, which
are denoted by $\delta\tau_r$. Assuming an analog signal which
requires a bandwidth $f_{bw}$ in order to resolve its
dynamics in time (i.e.\ a sampling time period $f_{bw}^{-1}$),
we measure $N = f_r / f_{bw}$ oscillations per sampled signal.
The relative accuracy of a measurement improves with $N$ as
\begin{equation}
  \sigma_r = \frac{\delta\tau_r}{\tau_r} \frac{1}{\sqrt{N}} =
    \frac{\delta\tau_r}{\tau_r} \sqrt{\frac{f_{bw}}{f_r}}\, .
  \label{eq:sigma_tau}
\end{equation}

Because bias current fluctuations $\delta I_b$ are linear to
frequency fluctuations $\delta f_r$ according to Eq.~(\ref{eq:df/dI}),
we can also conclude in first order that
$\delta\tau_r / \tau_r = \delta I_b / I_b$.
This yields an expression for the \textit{rms} current noise of the
signal sampled at $f_{bw}$:
\begin{equation}
  \left\langle \delta I_b \right\rangle_{f_{bw}} =
  \left\langle \delta I_b \right\rangle_{f_r} \sqrt{\frac{f_{bw}}{f_r}}
  = I_b \sigma_r\, .
\end{equation}
Consequently, in the case of random and uncorrelated $\delta\tau_r$
fluctuations we observe a white current noise spectrum with a density
\begin{equation}
  j_b = \frac{\left\langle \delta I_b \right\rangle_{f_{bw}}}{\sqrt{f_{bw}}} =
  \frac{I_b}{\sqrt{f_r}} \frac{\delta\tau_r}{\tau_r}\, ,
  \label{eq:I noise dens}
\end{equation}
apparently independent of $f_{bw}$.
Because phase (A) of a relaxation cycle dominates the timing,
we expect fluctuations in the critical current
$\delta I_c \propto \delta\tau_r$ to be
a major origin of $\delta\tau_r$ noise.

\subsection{Flicker $1/f$ noise in the critical current}
The critical current of Josephson junctions can fluctuate due to stochastic
charge trapping at defect sites in (or close to) the barrier, which are known
as ``two-level fluctuators''. A sufficiently large ensemble of such
fluctuators generates a $1/f$ spectrum, with significant contribution
only at low frequencies.
According to empirical models\cite{vanHarlingen2004,Wellstood2004}
the critical current noise density $j_\mathrm{tlf}$ due to $1/f$ flicker noise
can be described by
\begin{equation}
  j_{\mathrm{tlf}}^2 = \lambda\, \frac{I_c^2 T^2}{A f}\: ,
  \label{eq:tlf_noise_dens}
\end{equation}
where $A$ is the junction area and
$\lambda \approx 8.16 \times 10^{-24}\, \mathrm{m^2/K^2}$
is an average value obtained from collecting data over a wide range
of different junction parameters.\cite{vanHarlingen2004}
Scaling with $T^2$ was confirmed\cite{Wellstood2004} for
temperatures down to $90\, \mathrm{mK}$
(although the authors\cite{Wellstood2004} found a
higher noise level in their devices with
$\lambda \approx 39 \times 10^{-24}\, \mathrm{m^2/K^2}$).

\subsection{Critical current statistics from thermal activation
\label{Tactiv.sec}}
Escape from the zero-voltage state of a Josephson junction due to
thermal activation is a well-known and widely
studied phenomenon. It can be treated for a large variety of
junction types and external conditions. For our noise analysis
we can restrict ourselves to the simple ``transition-state''
model\cite{Kramers1940,Haenggi1990} where a particle inside a well is thermally
excited above the relative barrier potential and irreversibly leaves
the bound state.
This model is appropriate for underdamped Josephson junctions and is justified
by typical device parameters and experimental conditions as
given in the numerical examples (sections \ref{numeric.sec} and \ref{appl.sec}).
Particularly, we assume intermediate operation temperatures
satisfying
\begin{equation}
  \gamma = \frac{kT}{ E_J} \ll 1,
\end{equation}
where $E_J = \hbar I_c / 2e$ denotes a Josephson coupling energy,
sufficient to suppress the probability of retrapping from the running
state,\cite{Kivioja2005,Maennik2005,Krasnov2005}
but at the same time not too low to prevent macroscopic quantum
tunneling effects.
According to the model there is a nonvanishing probability for transitions
from the superconducting to the resistive state at current values $I_m < I_c$.
The lifetime of the zero-voltage state in a Josephson junction
as a function of the reduced current $i = I/I_c$ can be
expressed by\cite{Lee1971,Fulton1974}
\begin{equation}
  \tau_{\ell}^{-1}(i)=\frac{\omega_a}{2\pi} \, \mathrm{e}^{-U_0/kT}\, ,
  \label{eq:tau_ell}
\end{equation}
where $\omega_a = \omega_p (1-i^2)^{1/4}$ is the ``attempt frequency'' of the
particle in the well and
$U_0 = 2 E_J (\sqrt{1-i^2}- i \arccos i)$
is the relative potential height of the next barrier in the
Josephson junction ``washboard'' potential.
\footnote{%
The argument in the exponent of Eq.~(\ref{eq:tau_ell})
is in general notation $(-U_0/E_0)$ with
$E_0 \approx \max\left\{kT,\hbar\omega_a/2\pi\right\}$
in first approximation, taking into account the crossover
between classical and quantum limits. In the
range of interest ($\omega_a \rightarrow\! 0$ for $i \rightarrow\! 1$)
the excitation $E_0$ is dominated by $kT$.
}
The probability $P(t)$ for the junction to have switched from the
superconducting to the resistive state before time $t$ is\cite{Fulton1974}
\begin{equation}
  P(t)=1-\exp\left\{ -\int_{-\infty}^t \tau_{\ell}^{-1} I(t')\, \mathrm{d}t' \right\} .
  \label{eq:P(t)}
\end{equation}
By assuming small fluctuations compared to $I_c$ and using approximations
in the limit $\epsilon = 1 - i \ll 1$,
we can solve the integral by a similar approach to
Ref.~\onlinecite{Kurkijaervi1972} which yields
\begin{equation}
  P(\epsilon) = 1 - \exp\left\{ -\frac{\omega_p \gamma}{4\pi f_r (2\epsilon)^{1/4}}
              \exp\left(-\frac{2(2\epsilon)^{3/2}}{3\gamma}\right)\right\} .
  \label{eq:P(epsilon)}
\end{equation}
The mean value $\left\langle I_m \right\rangle $
of the observed critical current and its standard deviation
$\left\langle \delta I_m \right\rangle$
are calculated in the Appendix and are found to be
\begin{equation}
  \frac{\left\langle I_m \right\rangle}{I_c} \approx
   1-\frac{1}{5}\left(\gamma\ln\eta\right)^{2/3}
  \label{eq:Jm}
\end{equation}
and
\begin{equation}
  \frac{\left\langle \delta I_m \right\rangle}{I_c} \approx
   \frac{\gamma^{2/3}}{(\ln\eta)^{1/3}}
  \label{eq:dJm}
\end{equation}
where
\begin{equation}
  \eta=2\left(\frac{\omega_{p}}{2\pi f_{r}}\right)^6 \left(\frac{\gamma}{2}\right)^5 .
\end{equation}
Hence, $\left\langle \delta I_{m}\right\rangle $ is essentially
a function of $\gamma^{2/3}$, with a weak dependence on $(f_{r}/\omega_{p})$.
The approximations used for derivation of (\ref{eq:Jm},\ref{eq:dJm}) are
appropriate for $\eta\gg1$.
Similar results were obtained in Ref.~\onlinecite{Snigirev1983}.
The current noise density $j_m$ at the circuit input is, according
to Eq.~(\ref{eq:I noise dens}):
\begin{equation}
  j_m = \frac{I_b}{\sqrt{f_r}} \frac{\gamma^{2/3}}{(\ln\eta)^{1/3}}\, .
\end{equation}

To verify our results and to compare with other
models\cite{Carmeli1983,Barone1985} of different
formalism or treating different ranges of damping strength, we
evaluated numerically the probabilities $P(i)$, the transition
current distributions $p(i)=\mathrm{d}P(i)/\mathrm{d}i$,
and analyzed them with respect to shape, expectation
value $m_1$ and width $\sigma_m$. We found that, within
the range of allowed and reasonable model parameters,
only the mean values $m_1$ differed quantitatively
for different models, as should be expected for
different initial conditions and excitation forms.
However, there were insignificant differences in the distribution
shapes and particularly of their widths $\sigma_m$.
Therefore, Eq.~\ref{eq:dJm} can be considered a good estimate
for critical current fluctuations due to thermally activated
escape, applicable over a wide range of $\beta_c >1$.

\subsection{Other noise sources\label{noise_other.sec}}
Thermal current noise from ohmic resistors is dominated by
the shunt $R_s$ and corresponds to the standard Johnson noise
$j_s^2 = 4 kT / R_s$.
The voltage noise generated by $R_s$ and seen by the junction is,
due to  $f_r \propto V_s$, not amplified and
therefore equivalent to $j_s$ at the circuit input.

The real part of a good inductance $L$ vanishes. Therefore, $L$
can safely be considered as a ``cold resistor'' without thermal
noise contribution. Pickup of external magnetic noise can be shielded
and becomes negligible for small coils.

Because the Josephson current is a property of the ground state of the junction,
it does not fluctuate. Hence, shot noise in Josephson junctions is
only due to the quasiparticle current.
The relaxation oscillations within our concept are dominantly determined
by processes with the junction in the superconducting state.
Therefore, shot noise by itself should be negligible in our case.

However, as a consequence of the random nature of the junction phases in
the quasiparticle tunneling regime, the locking to the zero-voltage
state\cite{Fulton1971} at the end of an oscillation cycle
occurs within a time spread\cite{Adelerhof1994} on the order of
$ \delta \tau_z \approx 2 \pi \sqrt{L C_j} $. This results in an input
current noise density
\begin{equation}
  j_z = \frac{I_b}{\sqrt{f_r}} \frac{\delta \tau_z}{\tau_r}\, .
\end{equation}


\subsection{Noise conclusions\label{sub:Noise-conclusions}}
Combination of all noise sources derived above (and assumed
to be uncorrelated) yields a total circuit input current
noise density $j_b$ with
$j_b^2 = j_\mathrm{tlf}^2 + j_m^2 + j_s^2 + j_z^2$.
We make substitutions with respect to a notation
of $j_b$ in terms of the primary circuit and operation parameters
$L$, $R_s$, $R_n$, $I_c$, $\alpha$ and $T$:
\begin{equation}
  j_b^2 = c_1 \frac{(I_c R_n T)^2}{f} +
          c_2 \frac{\alpha I_c^{2/3} T^{4/3} L}{R_s} +
	  c_3 \frac{T}{R_s} +
	  c_4 \frac{\alpha^3 I_c^2 R_s}{R_n},
  \label{eq:jb_total2}
\end{equation}
where the constant coefficients $c_i$ are orthogonal
to the other parameters. Dependence  on junction
area $A$ and capacitance $C_j$ in Eq.~(\ref{eq:jb_total2})
is implicit by taking the products $R_n A = \rho_n$ and
$R_n C_j = \rho_n \epsilon_0 \epsilon_r /d$ to be constant
in standard Josephson junctions, respectively, where $\rho_n$
is the specific (normal) barrier resistance, $\epsilon_r$ the
barrier oxide dielectric constant and $d$ the barrier thickness.
Furthermore, we have neglected the $(\ln \eta)^{1/3}$ dependence
in Eq.~(\ref{eq:dJm}) assuming $\eta \gg 1$.
Hence, we can minimize the total circuit noise with
respect to the parameters in Eq.~(\ref{eq:jb_total2}).
In particular, $j_b$ appears to decrease with decreasing $T$,
$\alpha$ (although satisfying $\alpha \gg 1$) and $I_c$.
However, a lower $I_c$ has to be compensated by a larger
$L$ in order to satisfy Eq.~(\ref{eq:Lmin}), for the price
of lower $f_r \propto L^{-1}$ and a disadvantageous,
although weak increase of noise in the second term of
Eq.~(\ref{eq:jb_total2}).
Optimum values for $R_s$ and $R_n$ are found from a detailed
balance of the noise contributions. Assuming, for instance,
a negligible contribution from the fourth term in
(\ref{eq:jb_total2}), a large $R_s$ value seems favorable.
Due to $R_s \ll R_n$, however, we see a conflict with
a low noise requirement for the first term. This example
implies a not too large $R_n/R_s$ ratio.

\subsection{Dissipation and electron heating\label{sub:heating}}
Without formally relating dissipation processes 
to electronic noise, we consider local thermodynamics due to 
electron heating which may degrade the circuit's performance. 
The main current through the circuit is dissipated in the shunt 
resistor $R_s$ resulting in a permanent power of
$P_s = \alpha V_s I_c = R_s \alpha^2 I_c^2$. Successful removal 
of the excess energy is a matter of proper thermal anchoring 
(sufficiently large contact areas) to prevent
overheating of $R_s$. A similar power term arises during 
oscillations (in phases A and C) at the inductor $P_L = V_s I_c$.
Both these dissipation terms essentially constitute simple heat 
loads to the cryostat. Their magnitudes are typically very small
and the risk of local overheating of the lumped elements 
or excessive global heat overload is minor in standard 
cryogenic environments.

However, thermal nonequilibrium in the Josephson junction device 
may have significant consequences and deserves closer inspection.
Relevant dissipation in the junction can be shown to occur in 
phase B with  $P_j^B = V_g^2 C_j V_s / (2 I_c L)$ and in 
phase C with $P_j^C \approx V_s I_c/2$, where both are
(referred to a full oscillation cycle) weighted with their 
corresponding time constants. Satisfying Eq.~(\ref{eq:Cmax2}) yields
$P_j^B \ll V_s I_c/2$, hence the heat in the junction is essentially 
$P_j^C$. In order to remove the hot quasiparticles from 
the tunnel barrier region, normal metal trapping layers need to 
be attached adjacent to the superconducting junction electrodes.
At very low temperatures the interaction between
electronic and phonon subsystems in a normal metal vanishes, 
and the heat flow is described 
by $P_{ep} = \Sigma \nu (T_e^5 - T_p^5)$, where 
$\Sigma \approx 2.4\, \mathrm{nW}/(\mathrm{K}^5 \mu\mathrm{m}^3)$
is a material-dependent coupling constant,\cite{Wellstood1994} 
$\nu$ the volume of the normal metal and $T_e$ and $T_p$ are the 
temperatures of the electron and phonon system, respectively. 
The normal trapping layers will therefore experience an increased
electron temperature 
$T_e \approx (V_s I_c/(2 \Sigma \nu) + T_p^5)^{1/5}$.
Subsequently, the energy stored in the normal ``cooling fins'' has
to be transferred to the substrate. However, if the temperature
difference $| T_e - T_p |$ is small and the ``cooling fins'' are
made of thin films (i.e.\ large contact area to thickness ratio) 
the Kapitza resistance becomes negligible\cite{Wellstood1994} and
one can safely assume $T_p$ to be equal to the cryostat temperature.


\section{Numerical estimates\label{numeric.sec}}
In order to build a relaxation oscillation circuit we are
in principle free to choose any device or circuit
component and derive the remaining parameters based
on optimum arguments as discussed in the previous sections.
As an example we start with a Josephson junction of given
area $A$, junction (superconducting) material and an operation
temperature $T$.
The area determines $R_n = \rho_n / A$ with a typical
$\rho_n \approx 1\, \mathrm{k}\Omega\, \mu\mathrm{m}^2$ in our
standard devices and $C_j = \epsilon_0 \epsilon_r A/d$ with
$\epsilon_r \approx 8$ for AlO$_\mathrm{x}$ and
$d \approx 2\, \mathrm{nm}$. The choice of junction
material is a choice of energy gaps $\Delta_1$, $\Delta_2$,
determining $V_g$, $I_c^0$ and $E_J$. It is worth noting
that, according to Eq.~(\ref{eq:jb_total2}), a lower
$V_g \propto \kappa I_c R_n$ tends to result in a
lower noise level. A lower $V_g$ is also preferable to
exclude perturbations like Fiske modes from the gap
region. However, since $V_g$ delivers the
oscillation amplitude, a minimum level is required
for proper resolution of the oscillating signal
$V_\mathrm{out}$. This conflict can eventually
be alleviated instead by a suppression of $I_c^0$
by the factor $\kappa$ due to application of a
magnetic field, increasing the oscillation
frequency which may be a desired effect.
Finally, with the choice of operation point $\alpha$, the
required ratio $R_s/R_n \ll 1$ and a minimum $L$
satisfying (\ref{eq:Lmin}), the full properties of the
circuit are determined.

\begin{figure}
\includegraphics[width=84mm]{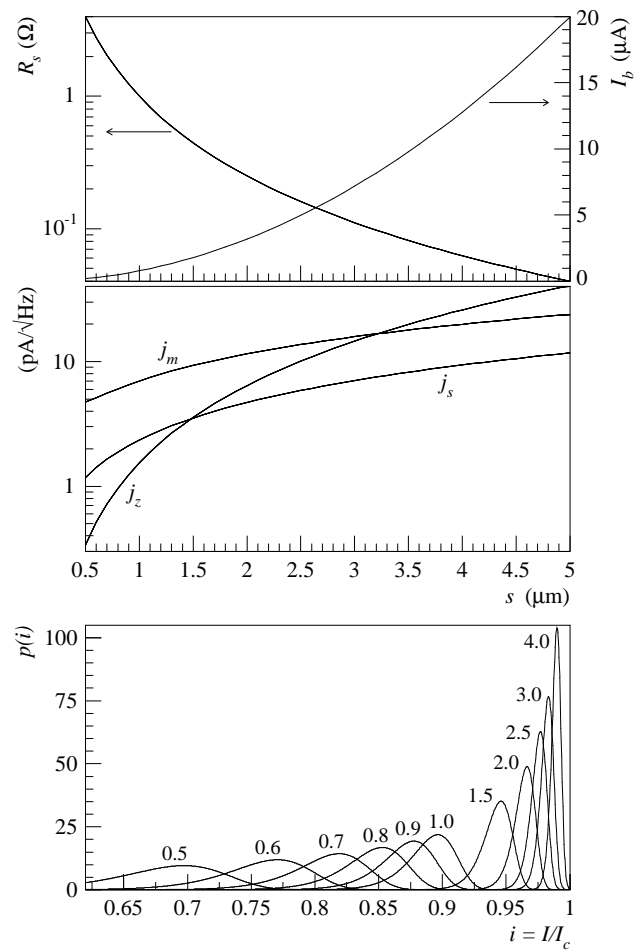}
\caption{\label{noise.f}%
Numerical calculations of $R_s$, $I_b$ and the different
contributions to circuit input current noise as a function
of junction side length $s$.
The corresponding transition
current distributions $p(i)=\mathrm{d}P(i)/\mathrm{d}i$
due to thermally activated zero-voltage escape are shown
in the bottom graph for different sizes $s$ (identified
by the numbers in units of $\mu\mathrm{m}$ adjacent to
the curves.)}
\end{figure}

Figure~\ref{noise.f} shows a numerical example of a
relaxation oscillation circuit as a function of
junction size (side length $s$), assuming an Aluminum junction
($\Delta_\mathrm{Al} = 170\, \mu\mathrm{V}$), an operation
temperature $T=100\, \mathrm{mK}$, the ratios $R_n/R_s=10^3$,
$I_c^0/I_c=1$ and $I_b/I_c=3$, and an effective $L$ which is
chosen 10 times larger than the minimum $L_\mathrm{min}$
in (\ref{eq:Lmin}).
The results in Fig.~\ref{noise.f} give an idea of the
order of the parameter ranges, including the contributions
from different noise sources. The flicker $1/f$ noise density
(at $f_r$) was always at least $10^4$ times lower than any
other noise contribution and is therefore not shown in
this example. The invariant parameters of this
configuration are:
$f_r = 5.23\, \mathrm{MHz}$,
$\omega_p = 151\, \mathrm{GHz}$,
$V_s = 0.80\, \mu\mathrm{V}$,
$\beta_c = 28.7$,
$\beta_s = \beta_c (R_s/R_n)^2 = 28.6 \times 10^{-6}$,
$\beta_L = 464.5$,
$\tau_{_B} = 45.1\, \mathrm{ps}$ and
$R_n C_j = 35.4\, \mathrm{ps}$.

It is apparent in Fig.~\ref{noise.f} that fluctuations
due to thermally activated zero-voltage escape
(i.e.\ current noise $j_m$) are the dominant noise
process in the range of small devices.
For illustration the corresponding distribution functions
$p(i)=\mathrm{d}P(i)/\mathrm{d}i$ of transition currents
are included in Fig.~\ref{noise.f}. In spite of increasing
distribution width with decreasing junction size $s$, the
noise density decreases due to a faster decay of $I_b$.

\begingroup
\squeezetable
\begin{table*}
\caption{\label{table1.t}%
List of component and operation parameter values of a
relaxation oscillation circuit optimized with respect
to minimum input current noise. Initial conditions
are described in the main text.}
\begin{ruledtabular}
\begin{tabular}{cccccccccccccccccccc}
$A$ & $R_n$ & $R_s$ & $C_j$ & $I_c$ & $I_b$ & $V_s$ & $L$ & $\beta_c$
    & $\beta_L$ & $f_r$ & $\omega_p$ & $\tau_{_B}$ & $\delta\tau_z$
    & $I_m/I_c$ & $\delta I_m/I_c$ & $j_\mathrm{tlf}$ & $j_m$ & $j_s$ & $j_z$
    \\
$(\mu\mathrm{m}^2)$ & $(\Omega)$ & $(\Omega)$ & (fF) & (nA) & (nA) & $(\mu\mathrm{V})$
    & $(\mu\mathrm{H})$ & & & (MHz) & (GHz) & (ps) & (ns) & &
    & \multicolumn{4}{c}{$(\mathrm{fA}/\sqrt{\mathrm{Hz}})$\footnote{%
    These units refer to all four current noise density terms.}}
    \\ \hline
0.1 & $10^4$ & $10^2$ & 3.54 & 25.0 & 75.1 & 7.51 & 2.87 & 26.9
    & 218 & 104.5 & 146.6 & 44.9 & 0.63 & 0.56 & 0.098
    & 0.002\footnote{Value taken at $f_r$.} & 965 & 235 & 363
    \\
\end{tabular}
\end{ruledtabular}
\end{table*}
\endgroup

As a second numerical example we calculate a realistic
minimum-noise circuit configuration without leaving the
range of classical dynamics of Josephson junctions
as assumed for our model.
We choose an Al/AlO$_\mathrm{x}$/Al junction where we
expect the best quality tunnel barriers and a sufficient
oscillation output amplitude
$V_\mathrm{out} \approx 340\, \mu\mathrm{V}$.
The minimum junction size is restricted by the range of
validity of our model, requiring
$E_J > E_C = e^2 / 2 C_j$ and $E_J > kT$ to prevent
single-electron charging or macroscopic quantum
tunneling effects, respectively.\footnote{%
The phase diffusion model in Ref.~\onlinecite{Kautz1990}
does not significantly alter the switching behavior (and
thereby the noise) of our dynamical circuit even for
moderate $E_J/E_C$ ratios, as long as $\beta_c$ is
sufficiently large.}
This is just satisfied with a junction of area
$A = 0.1\, \mu \mathrm{m}^2 \approx (316\, \mathrm{nm})^2$,
which can be fabricated by standard $e$-beam lithography.
An operation temperature of $T = 100\, \mathrm{mK}$ is
easily reached and maintained in modern cryostats
even in the case of some moderately low dissipation
in the circuit.
The values for the circuit components follow from our
definitions of
$\alpha=3$,
$R_n/R_s=100$,
$\kappa=1$,
$L/L_\mathrm{min}=5$,
and are listed in Table~\ref{table1.t}.
The results show a total input current noise as low as
about $1\, \mathrm{pA}/\sqrt{\mathrm{Hz}}$, with the
dominant contribution from thermally activated zero-voltage
escape. Flicker $1/f$ noise density at $f_r$ is at a
negligible level and remains insignificant
down to very low readout bandwidths $f_{bw}$.
The total noise figure of this configuration is
well competitive with the best commercial
SQUID amplifiers. In addition, due to the advantage
of improving noise behavior with increasing oscillation
frequency, it delivers a bandwidth superior to most
SQUID systems.

It is clear that the operation point for the current
or voltage biased device under test is fixed to
$I_b$ or $V_s$ in this example. For
devices requiring different bias values (as in
Ref.~\onlinecite{FurlanRelax}) the circuit
components have to be adapted with respect to
the specifications.

As discussed in Sec.~\ref{sub:heating} the total 
dissipated power
$P_\mathrm{tot} \approx V_s I_c (\alpha + 3/2)$
in the circuit example above (Table~\ref{table1.t}) amounts
to about $0.87\, \mathrm{pW}$. We assume the normal metal 
``cooling fins'' attached to the junction electrodes
to be thin films of $100\, \mathrm{nm}$ thickness and 
with an area of $10 \times 10\, \mu\mathrm{m}^2$. The 
effective electron temperature in these normal metal traps
will then increase (from $100\, \mathrm{mK}$) to about
$108\, \mathrm{mK}$, which is absolutely acceptable 
with respect to power equilibration in the system 
as well as to the thermal properties and operation 
of the Josephson junction. Larger tunnel junctions will
require proportionally enlarged ``cooling fin'' areas, 
which is feasible up to a practical size compared
to the bulky dimensions of $L$ (and eventually $R_s$).
Furthermore, if the normal metal traps also form the 
connecting leads, the trapping and cooling efficiencies
may improve appreciably.


\section{Possible applications\label{appl.sec}}

We have developed the relaxation oscillation
analog-to-frequency converter primarily for readout
of cryogenic radiation detectors.\cite{Booth1996}
The aim was to overcome problems or limitations in
scaling to large number pixel readout.
Besides the outstanding noise properties, a
particularly nice feature of the relaxation oscillation
circuit is its potential for simple implementation into a
frequency-domain multiplexing scheme by tuning
the oscillation frequencies of the individual
analog-to-frequency converters to well separated
values, and then using one single line to read them out.
It should be taken into account, however,
that a signal excursion from a detector generates
a frequency shift, which should not overlap with a
neighboring oscillator in the simplest case.
A more sophisticated scheme could lock into the
``dark'' (no detector signal) characteristic
frequencies and, upon disappearance of one channel due
to an analog detector signal, remove the other frequency
bands in order to recover the signal of interest.\footnote{%
In case of frequency band overlap, the signal, which
is only partially recovered, can be reconstructed from
a decent knowledge of the expected pulse shape.}
The circuit example in Table~\ref{table1.t} is apparently
not a good choice for a multiplexing readout due to fairly
broad $\delta I_m / I_c$. However, if we accept a moderate
increase in noise level by choosing larger junctions, 
fluctuations are easily reduced to 
$\delta I_m / I_c < 0.01$ (see Fig.~\ref{noise.f}). 
In that case, and taking into account that the circuits
should all operate in a bias range $V_s \lesssim V_g /3$,
we estimate that up to about 10 oscillators can be implemented
with sufficient separation. The different frequencies can
be tuned by either different shunt resistors $R_s$,
different inductors $L$ or a careful variation of 
junction sizes  (resulting in different $I_c$), 
whatever best meets experimental conditions and 
technical possibilities. 
One should also understand that combination of  
several circuit outputs by simple connection (e.g.\
through resistors, no amplification) reduces the amplitudes
of the individual signals by a factor equal
to the number of interconnecting channels. Required 
minimum signal-to-noise therefore limits the 
feasible number of multiplexed channels.       

\begin{figure}
\includegraphics[width=84mm]{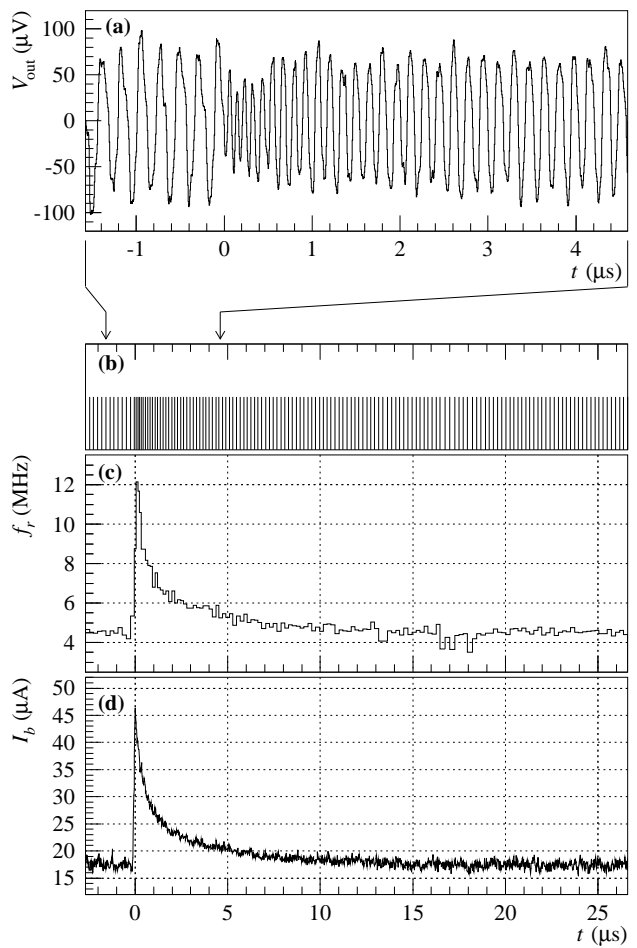}
\caption{\label{signal.f}%
(a)~Response of the relaxation oscillation circuit
modulated due to a $6\, \mathrm{keV}$ X-ray event
in a SINIS detector at $80\, \mathrm{mK}$.
The sinusoidal shape and reduced amplitude of the
oscillations are due to band-pass filtering at the
post-amplification stage. (b)~Rectified oscillator
signal where each line represents the position of an
oscillation cycle maximum (note the larger time scale,
applying to all three lower plots).
(c)~Inverse oscillation periods as a function of
time, equivalent to a time-dependent frequency $f_r$.
(d)~Analog signal from the same detector measured
with conventional electronics and taken at
corresponding experimental conditions.}
\end{figure}

To test and demonstrate the working principle of a 
(single) relaxation oscillation circuit readout
we have measured the response of a SINIS
microcalorimeter\cite{FurlanSINIS} to irradiation
with $6\, \mathrm{keV}$ X-rays.
The detector which replaced $R_b$ was voltage biased.
Figure~\ref{signal.f} shows the results of an X-ray
event. The circuit and device parameters were:
$L=48\, \mathrm{nH}$,
$R_s=91\, \mathrm{m}\Omega$,
junction size $s=15\, \mu\mathrm{m}$ and effective
critical current $I_c=7.28\, \mu\mathrm{A}$ $(\kappa=8)$.
The detector's ``dark'' (or bias) current was
$I_b^0 = 17.5\, \mu\mathrm{A} = 2.4 I_c$, the measured
analog signal peak current was
$I_b^1 = 46.4\, \mu\mathrm{A}$, as shown in
Fig.~\ref{signal.f}d.
The relaxation oscillation frequencies from
Fig.~\ref{signal.f}c, taken at the same
operation point and conditions, were
$f_r^0 = 4.47\, \mathrm{MHz}$ and
$f_r^1 = 12.1\, \mathrm{MHz}$, respectively.
Taking into account the conversion factor
$\mathrm{d}f_r/\mathrm{d}I_b = 260\, \mathrm{kHz}/\mu\mathrm{A}$,
see Eq.~(\ref{eq:df/dI}), the analog and the frequency-modulated
signal are perfectly compatible quantitatively as well as
qualitatively (pulse shape). The noise level is about the same
in both cases and is due to  detector noise.
The circuit noise alone is estimated to contribute about
$0.3\, \mu\mathrm{A}$ {\em rms} integrated over full
bandwidth up to $f_r$.
We should say that the microcalorimeter device and circuit
configuration are by no means optimal in this example,
they are rather a preliminary choice of available components.
Primarily, these results are of illustrative nature, demonstrating
the principle and feasibility of cryogenic detector readout.

Other possible applications for relaxation oscillation based
analog-to-frequency conversion can be, in a wider sense, considered
for any type of cryogenic device operated at relatively low bias
levels, exhibiting small variations of its electronic properties
or actively delivering small analog signals.
A list may include quantum dots and wires,
single-electron devices and quantum Hall structures, to name a few.
Due to the large bandwidth, the readout method is also attractive
for detection of fast processes like quantum noise or background
charge fluctuations.
The resistors $R_b$ and $R_s$ in our scheme just represent a
current and a voltage source, respectively, and can be replaced
by the device of choice.

It is important to note, however, that the oscillator junction
characteristics (essentially represented by $I_c$) may slightly
vary from cooldown to cooldown and therefore cause a measurable
spread in the conversion factor~(\ref{eq:df/dI}). Therefore, the
circuit is unfortunately not appropriate for absolute measurements
on a level as required e.g.\ by metrologists.

As a concluding experiment we propose a setup for high-precision
thermometry at low temperatures, replacing the classical
four-point measurement on thermistors. The temperature-sensitive
element would typically replace $R_s$ to minimize dissipation.
The difficulty of applying small analog excitations and detecting
low output levels (across long wires), competing with noise, is
circumvented by directly ``digitizing'' the small signal very
close to the sensor with a low-noise converter. It is clear that
this thermometer readout can only be operated in a limited
temperature range (presumably one order of magnitude) where the
junction dynamics (fluctuations, switching probabilities) are
sufficiently insensitive to $T$ variations.


\section{Conclusions\label{conclus.sec}}
We have investigated the feasibility
of a cryogenic low-noise analog-to-frequency converter
with acceptable linearity over a broad range of
circuit and operation parameters. The dynamical
behavior can be well described by simple circuit
theory and classical models of the single Josephson
junction involved. Their agreement with
experimental data is perfect.
We have presented a thorough analysis of noise sources,
where fluctuation processes in the Josephson junction
appear to usually dominate the circuit's noise figure for
typical configurations and experimental conditions.
The inherent broadband operation
paired with very good noise performance offers a
versatile system for a wide range of applications.
As one possible example we have demonstrated the
readout of a cryogenic microcalorimeter exposed
to X-rays. Implementation into a multiplexing
scheme was discussed and needs to be experimentally
tested for a large channel-number readout.


\begin{acknowledgments}
We are grateful to Eugenie Kirk for excellent device fabrication,
to Philippe Lerch and Alex Zehnder for valuable and stimulating
discussions, and to Fritz Burri for technical support.
\end{acknowledgments}


\appendix*
\section{}
For the analysis of noise due to thermally activated zero-voltage
escape an evaluation of the central moments of the probability
density function $p(i)=\mathrm{d}P(i)/\mathrm{d}i$ is required,
yielding expectation value $m_1 = \int i\, p(i)\, \mathrm{d}i$
and variance $m_2 = \int (i-m_1)^2 p(i)\, \mathrm{d}i$, where
$P(i)$ is given by Eq.~(\ref{eq:P(epsilon)}) and
$\sigma_m = \sqrt{m_2}$ subsequently denotes the standard deviation.
Analytical integration can be circumvented, however,
by approximating  $p(i)$ with a Gaussian distribution
and solving for the appropriate values satisfying
$P(m_1)=0.5$ and
$P(m_1\pm\sigma_m) = 0.5 \{ 1 \pm \mathrm{erf} (1/\sqrt{2})\}$,
respectively. A general solution of $P(m)=h$ is
\begin{equation}
  m = \frac{1}{8} \left\{2\gamma W
  \left(\frac{\eta}{\ln(1-h)^{6}}\right)\right\}^{2/3},
  \label{eq:m_general}
\end{equation}
where $\eta=2(\omega_p / 2\pi f_r)^6 (\gamma/2)^5$
and  $W$ is the Lambert function satisfying
$W(x)\cdot\exp\left(W(x)\right)=x$.
An asymptotic expansion of $W$ for $(x\rightarrow\infty)$ can be written as
$W(x)  =  \ln x-\ln(\ln x)$,
where higher order terms are suppressed.
Setting $h=0.5$ in Eq.~(\ref{eq:m_general}) yields the expectation
value
\[ \frac{\left\langle I_m \right\rangle}{I_c} = m_1 \approx
  1-\frac{1}{8} \left(2\gamma \ln \frac{9\eta}{\ln(9\eta)}\right)^{2/3}. \]
In the limit $(\eta\rightarrow\infty)$ this expression approaches
\begin{equation}
  \frac{\left\langle I_m \right\rangle}{I_c} \approx
  1-\frac{1}{8}\left(2\gamma\ln\eta\right)^{2/3}.
\end{equation}
Correspondingly, for $\left\langle\delta I_m \right\rangle /I_c = \sigma_m$
we find an expansion
\[ \frac{\sigma_m}{\frac{1}{8}\left(2\gamma\right)^{2/3}}
  \approx \frac{4.73}{(\ln\eta)^{1/3}} +
  \frac{1.58\ln\left(\ln\eta\right)-10.15}{(\ln\eta)^{4/3}} , \]
which, for $(\eta\rightarrow\infty)$, approaches
\begin{eqnarray}
  \frac{\left\langle\delta I_m \right\rangle }{I_c}
   & \approx & \frac{1}{8}\left(2\gamma\right)^{2/3}\frac{4.73}{(\ln\eta)^{1/3}}
   \nonumber \\
   & \approx & 0.94 \frac{\gamma^{2/3}}{(\ln\eta)^{1/3}}\,  .
\end{eqnarray}
If $m_1$ significantly deviates from unity the approximation
\begin{equation}
  \frac{\left\langle\delta I_m \right\rangle }{\left\langle I_m \right\rangle }
  \approx \frac{\gamma^{2/3}}{(\ln\eta)^{1/3}} +
  \frac{3}{16} \gamma^{4/3} (\ln\eta)^{1/3} +
  \mathrm{O} \left( \gamma^2 \ln\eta \right)
\end{equation}
describes the relative distribution width of transition current fluctuations.

\end{document}